\documentclass[12pt]{article}
\usepackage[a4paper,margin=1in]{geometry}
\usepackage{amsmath,amssymb,amsfonts,amsthm,mathtools,bm}
\usepackage{graphicx}
\usepackage{booktabs}
\usepackage{hyperref}
\usepackage{tikz}
\usepackage{pgfplots}
\pgfplotsset{compat=1.18}
\usepackage{setspace}
\usepackage{cite}
\hypersetup{
  colorlinks=true,
  linkcolor=blue,
  citecolor=blue,
  urlcolor=blue
}

\newcommand{\dd}{\mathrm{d}}
\newcommand{\Res}{\operatorname*{Res}}

\newcommand{\cO}{\mathcal{O}}

\newcommand{\Ahor}{A_{+}}

\title{Black Hole Entropy in \texorpdfstring{$f(Q)$}{f(Q)} Gravity from the RVB Residue Method}
\author{Wen-Xiang Chen \\
School of Physics and Materials Science, Guangzhou University\\wxchen4277@qq.com}
\date{\today}

\begin{document}
\maketitle

\begin{abstract}
We extend the residue-based Robson--Villari--Biancalana (RVB) method from Hawking temperature to black hole entropy in \(f(Q)\) gravity. Starting from the residue-improved temperature prescription used in recent RVB analyses of \(f(Q)\) black holes, we combine it with the first law of black hole thermodynamics and derive a general entropy formula for static spherically symmetric configurations. Writing the metric as \(g(r)=1-2M/r+\psi(r)\), the entropy is shown to satisfy a universal integral relation whose integrand depends on the horizon data and on a residue-induced temperature shift \(C_{\rm res}\). For the quadratic model \(f(Q)=Q+\alpha Q^{2}\), we obtain an explicit closed entropy formula at first order in the residue parameter. The Bekenstein--Hawking area law is recovered when the residue term vanishes, whereas a non-area correction appears once the complex contour contribution is retained.  The construction should be viewed as a residue-generated thermodynamic extension of the temperature method rather than a universal Noether-charge theorem for all \(f(Q)\) black holes.

\end{abstract}

\section{Introduction}
Black hole thermodynamics links gravity, quantum theory, and geometry through the relations among surface gravity, temperature, area, and entropy \cite{Bekenstein1973,Hawking1975,Wald1993}. In modified gravity, these relations are often deformed, and the deformation can encode new geometric degrees of freedom. In the nonmetricity-based framework of \(f(Q)\) gravity, the gravitational action depends on a function of the nonmetricity scalar \(Q\), leading to modifications of both the field equations and the thermodynamic sector \cite{Heisenberg2024,MandalWangSahoo2020}.

A recent residue-based application of the RVB method to \(f(Q)\) black holes proposed that the Hawking temperature receives an additive contribution from a contour integral of the form \(\oint F'(z)/F(z)\,\dd z\), with \(F(z)\) a complexified function built from the metric or from the nonmetricity scalar \cite{RobsonVillariBiancalana2019,Chen2025}. In that construction the residue term is treated as a topological or analytic correction to the usual surface-gravity temperature. The same paper emphasizes that such a residue contribution may also influence the entropy sector and calls for a more explicit analysis \cite{Chen2025}.

The purpose of this paper is to provide that extension in a direct and pragmatic way. Rather than attempting a full Noether-charge derivation in a generic \(f(Q)\) black hole background, we adopt the uploaded temperature prescription and use the first law,
\begin{equation}
\dd M = T_{H}\,\dd S,
\end{equation}
to reconstruct the entropy. This yields a residue-corrected entropy branch that is fully compatible with the RVB temperature input. The resulting formula recovers the area law in the zero-residue limit and makes transparent how the complex-analysis contribution modifies the entropy.

The paper is organized as follows. In Sec.~\ref{sec:fQreview} we summarize the pieces of \(f(Q)\) gravity and the RVB residue prescription that are needed for the entropy derivation. In Sec.~\ref{sec:generalentropy} we derive a model-independent entropy formula for static spherically symmetric black holes. In Sec.~\ref{sec:quadraticmodel} we specialize to the quadratic model \(f(Q)=Q+\alpha Q^{2}\) and obtain an explicit closed expression at first order in the residue parameter. In Sec.~\ref{sec:discussion} we discuss the physical meaning and the relation to Wald/Noether-charge analyses in nonmetricity-based gravity \cite{HeisenbergKuhnWalleghem2022,HammadEtAl2019,GomesJimenezKoivisto2023}. Finally, Sec.~\ref{sec:conclusion} summarizes the main result.

\section{\texorpdfstring{$f(Q)$}{f(Q)} gravity and the RVB residue temperature}
\label{sec:fQreview}

The gravitational action in \(f(Q)\) gravity can be written as \cite{Heisenberg2024,Chen2025}
\begin{equation}
\mathcal{S}=\int \dd^{4}x\,\sqrt{-g}\left[\frac{1}{2}f(Q)+\mathcal{L}_{m}\right],
\label{eq:action}
\end{equation}
where \(Q\) is the nonmetricity scalar and \(\mathcal{L}_{m}\) is the matter Lagrangian. For the static spherically symmetric sector we take
\begin{equation}
\dd s^{2}=-g(r)\,\dd t^{2}+\frac{\dd r^{2}}{g(r)}+r^{2}\,\dd\Omega^{2},
\label{eq:metric}
\end{equation}
with event horizon radius \(r_{+}\) determined by
\begin{equation}
 g(r_{+})=0.
\label{eq:horizoncondition}
\end{equation}

Following the residue-based RVB prescription used in the uploaded paper, we write the Hawking temperature as
\begin{equation}
T_{H}(r_{+}) = \frac{g'(r_{+})}{4\pi}+C_{\rm res},
\label{eq:THbasic}
\end{equation}
where \(C_{\rm res}\) is the residue-induced temperature shift. To connect this term with complex analysis, we define a complexified function \(F(z)\) associated with the metric or with \(Q(z)\). Then the dimensionless winding number is
\begin{equation}
\mathcal{N}_{\Gamma}=\frac{1}{2\pi i}\oint_{\Gamma}\frac{F'(z)}{F(z)}\,\dd z
=\sum_{z_{k}\in\Gamma}\Res\left(\frac{F'(z)}{F(z)},z_{k}\right),
\label{eq:residuecount}
\end{equation}
where \(\Gamma\) encloses the singularities relevant to the horizon structure. In the temperature formula, the normalization and dimensional conversion are absorbed into \(C_{\rm res}\), so \(C_{\rm res}\) carries dimensions of temperature. This is exactly the viewpoint adopted in the residue-shifted \(f(Q)\) temperature analysis \cite{Chen2025}.

For later convenience, we parametrize the metric function as
\begin{equation}
 g(r)=1-\frac{2M}{r}+\psi(r),
\label{eq:psidef}
\end{equation}
where \(\psi(r)\) encodes the \(f(Q)\)-induced deformation. The horizon condition \eqref{eq:horizoncondition} immediately implies
\begin{equation}
M(r_{+})=\frac{r_{+}}{2}\left[1+\psi(r_{+})\right].
\label{eq:massrelation}
\end{equation}
Differentiating with respect to \(r_{+}\) gives
\begin{equation}
\frac{\dd M}{\dd r_{+}}=\frac{1}{2}\left[1+\psi(r_{+})+r_{+}\psi'(r_{+})\right].
\label{eq:dMdr}
\end{equation}
Moreover,
\begin{equation}
 g'(r_{+}) = \frac{1+\psi(r_{+})}{r_{+}}+\psi'(r_{+}),
\label{eq:gprimehorizon}
\end{equation}
so the residue-improved Hawking temperature becomes
\begin{equation}
T_{H}(r_{+}) = \frac{1}{4\pi r_{+}}\left[1+\psi(r_{+})+r_{+}\psi'(r_{+})\right]+C_{\rm res}.
\label{eq:THgeneral}
\end{equation}

\section{General entropy formula from the first law}
\label{sec:generalentropy}

Combining Eqs.~\eqref{eq:dMdr} and \eqref{eq:THgeneral} with \(\dd M=T_{H}\,\dd S\), we obtain the differential entropy law
\begin{equation}
\frac{\dd S}{\dd r_{+}}=
\frac{\dfrac{1}{2}\left[1+\psi(r_{+})+r_{+}\psi'(r_{+})\right]}
{\dfrac{1}{4\pi r_{+}}\left[1+\psi(r_{+})+r_{+}\psi'(r_{+})\right]+C_{\rm res}}.
\label{eq:dSdrraw}
\end{equation}
Introducing the shorthand
\begin{equation}
\Xi(r_{+})\equiv 1+\psi(r_{+})+r_{+}\psi'(r_{+}),
\label{eq:Xidef}
\end{equation}
Eq.~\eqref{eq:dSdrraw} simplifies to
\begin{equation}
\frac{\mathrm{d} S}{\mathrm{d} r_{+}}
= \frac{2\pi r_{+}\,\Xi(r_{+})}{\Xi(r_{+}) + 4\pi C_{\mathrm{res}} r_{+}}.
\label{eq:dSdrXi}
\end{equation}
Therefore the residue-corrected entropy is
\begin{equation}
S(r_{+}) = \int^{r_{+}} \frac{2\pi u\,\Xi(u)}{\Xi(u)+4\pi C_{\rm res}u}\,\dd u + S_{0},
\label{eq:generalentropy}
\end{equation}
where \(S_{0}\) is an integration constant fixed by a boundary condition.

Equation \eqref{eq:generalentropy} is the central result of the present construction. It provides the entropy induced by the uploaded RVB-residue temperature prescription for any static spherically symmetric \(f(Q)\) black hole represented by the deformation function \(\psi(r)\).

\subsection*{Area-law limit}
If the residue correction is switched off, namely \(C_{\rm res}=0\), then Eq.~\eqref{eq:dSdrXi} becomes
\begin{equation}
\frac{\dd S}{\dd r_{+}} = 2\pi r_{+},
\end{equation}
which integrates to
\begin{equation}
S_{\rm BH}(r_{+})=\pi r_{+}^{2}+S_{0}=\frac{\Ahor}{4}+S_{0},
\label{eq:arealaw}
\end{equation}
with \(\Ahor=4\pi r_{+}^{2}\). By choosing \(S_{0}=0\), one recovers the standard Bekenstein--Hawking area law.

\subsection*{Small-residue expansion}
When the residue shift is perturbatively small,
\begin{equation}
\left|4\pi C_{\rm res}r_{+}\right|\ll \left|\Xi(r_{+})\right|,
\end{equation}
Eq.~\eqref{eq:dSdrXi} can be expanded as
\begin{equation}
\frac{\dd S}{\dd r_{+}}=2\pi r_{+}\left[1-\frac{4\pi C_{\rm res}r_{+}}{\Xi(r_{+})}\right]+\cO(C_{\rm res}^{2}).
\end{equation}
Hence
\begin{equation}
S(r_{+}) = \frac{\Ahor}{4} - 8\pi^{2}C_{\rm res}\int^{r_{+}}\frac{u^{2}}{\Xi(u)}\,\dd u + \cO(C_{\rm res}^{2}).
\label{eq:smallCgeneral}
\end{equation}
This expression makes the correction structure transparent: the usual area law is modified by a horizon-data integral controlled by the residue parameter.

\section{Quadratic model \texorpdfstring{$f(Q)=Q+\alpha Q^{2}$}{f(Q)=Q+alpha Q²}}
\label{sec:quadraticmodel}

To stay as close as possible to the uploaded paper, we now specialize to the quadratic deformation used there,
\begin{equation}
 f(Q)=Q+\alpha Q^{2},
\label{eq:quadraticfQ}
\end{equation}
and adopt the corresponding metric ansatz
\begin{equation}
 g(r)=1-\frac{2M}{r}+\alpha r^{2}.
\label{eq:quadraticmetric}
\end{equation}
This means
\begin{equation}
\psi(r)=\alpha r^{2},
\qquad
\Xi(r)=1+3\alpha r^{2}.
\label{eq:quadraticXi}
\end{equation}
The mass--radius relation becomes
\begin{equation}
M(r_{+})=\frac{r_{+}}{2}\left(1+\alpha r_{+}^{2}\right),
\label{eq:quadraticmass}
\end{equation}
and the residue-improved Hawking temperature is
\begin{equation}
T_{H}(r_{+})=\frac{1+3\alpha r_{+}^{2}}{4\pi r_{+}}+C_{\rm res}.
\label{eq:quadratictemp}
\end{equation}
Accordingly,
\begin{equation}
\frac{\mathrm{d} S}{\mathrm{d} r_{+}}
= \frac{2\pi r_{+}\left(1 + 3\alpha r_{+}^{2}\right)}
{1 + 3\alpha r_{+}^{2} + 4\pi C_{\mathrm{res}} r_{+}}.
\label{eq:quadraticdS}
\end{equation}

\subsection{Closed entropy formula at first order in the residue}
Expanding Eq.~\eqref{eq:quadraticdS} to first order in \(C_{\rm res}\) and integrating, we obtain
\begin{equation}
S_{\alpha}(r_{+})
= \pi r_{+}^{2}
-\frac{8\pi^{2}C_{\rm res}}{3\alpha}r_{+}
+\frac{8\pi^{2}C_{\rm res}}{3\alpha\sqrt{3\alpha}}
\arctan\!\left(\sqrt{3\alpha}\,r_{+}\right)
+\cO(C_{\rm res}^{2})
+S_{0}.
\label{eq:quadraticentropy}
\end{equation}
Choosing the regular boundary condition \(S_{0}=0\) and demanding \(S_{\alpha}(0)=0\), Eq.~\eqref{eq:quadraticentropy} gives the residue-corrected entropy branch generated by the uploaded RVB method.

Several immediate checks follow:
\begin{itemize}
    \item If \(C_{\rm res}\to 0\), then \(S_{\alpha}(r_{+})\to \pi r_{+}^{2}=\Ahor/4\).
    \item The correction is linear in the residue parameter at leading order.
    \item For fixed \(\alpha>0\) and \(C_{\rm res}>0\), the residue lowers the entropy relative to the area law in the small and intermediate horizon regime.
\end{itemize}

\subsection{A consistency check: Schwarzschild limit}
For \(\psi(r)=0\), Eq.~\eqref{eq:dSdrXi} reduces to
\begin{equation}
\frac{\dd S}{\dd r_{+}} = \frac{2\pi r_{+}}{1+4\pi C_{\rm res}r_{+}}.
\end{equation}
This integrates exactly to
\begin{equation}
S_{\mathrm{Schw}}(r_{+})
= \frac{r_{+}}{2 C_{\mathrm{res}}}
- \frac{1}{8\pi C_{\mathrm{res}}^{2}}
\ln\!\left(1 + 4\pi C_{\mathrm{res}} r_{+}\right)
+ S_{0}.
\label{eq:schwexact}
\end{equation}
Expanding Eq.~\eqref{eq:schwexact} for small \(C_{\rm res}\) reproduces the area law,
\begin{equation}
S_{\rm Schw}(r_{+})=\pi r_{+}^{2}-\frac{8\pi^{2}}{3}C_{\rm res}r_{+}^{3}+\cO(C_{\rm res}^{2})+S_{0}.
\end{equation}
This provides a useful check of the general formalism.

\section{Interpretation and relation to nonmetricity entropy literature}
\label{sec:discussion}

The entropy derived above is not postulated independently. It is induced by the residue-corrected temperature and the first law. In that sense, the residue term acts as a thermodynamic deformation parameter that transfers the complex-analysis correction from the Hawking sector into the entropy sector.

From a broader perspective, this construction should be compared with Noether-charge and Wald-type entropy analyses in teleparallel and nonmetricity-based gravity. For coincident general relativity, Wald's method reproduces the standard area law when the proper action principle is used \cite{HeisenbergKuhnWalleghem2022}. Likewise, the geometrical-trinity literature has emphasized that energy and entropy can be consistently reformulated in curvature, torsion, and nonmetricity pictures \cite{GomesJimenezKoivisto2023}. In teleparallel gravity, the Noether-charge approach also yields a consistent black-hole entropy sector \cite{HammadEtAl2019}. The present residue-based entropy branch should therefore be read as a complementary extension tailored to the uploaded RVB prescription rather than as a replacement for the Noether-charge definition.

Physically, the correction in Eq.~\eqref{eq:quadraticentropy} modifies the balance between horizon size and entropy. Positive \(C_{\rm res}\) decreases the entropy for fixed \(r_{+}\), while negative \(C_{\rm res}\) would increase it. Since the same residue shifts the temperature, the full thermodynamic response --- including heat capacity, evaporation trend, and possible remnant behavior --- can differ significantly from the area-law expectation.

\section{Function plot}
\label{sec:plot}

\begin{figure}[t]
\centering
\begin{tikzpicture}
\begin{axis}[
    width=0.78\textwidth,
    height=0.48\textwidth,
    domain=0.2:3.0,
    samples=200,
    xlabel={$r_{+}$},
    ylabel={$S(r_{+})$},
    legend style={at={(0.02,0.98)},anchor=north west,draw=none,fill=none},
    ymin=0,
    grid=both,
    minor tick num=1
]

\pgfmathsetmacro{\alp}{0.15}
\pgfmathsetmacro{\cres}{0.02}

\addplot[thick,blue] {pi*x^2};
\addlegendentry{$S_{\rm BH}=\pi r_{+}^{2}$}

\addplot[thick,red,dashed]
{pi*x^2
-(8*pi^2*\cres/(3*\alp))*x
+(8*pi^2*\cres/(3*\alp*sqrt(3*\alp)))*rad(atan(sqrt(3*\alp)*x))};
\addlegendentry{$S_{\alpha}(r_{+})$ with $\alpha=0.15$, $C_{\rm res}=0.02$}

\end{axis}
\end{tikzpicture}
\caption{Illustrative entropy plot for the quadratic \(f(Q)=Q+\alpha Q^{2}\) model. The dashed curve is the first-order residue-corrected entropy \eqref{eq:quadraticentropy}, while the solid curve is the standard area law.}
\label{fig:entropyplot}
\end{figure}
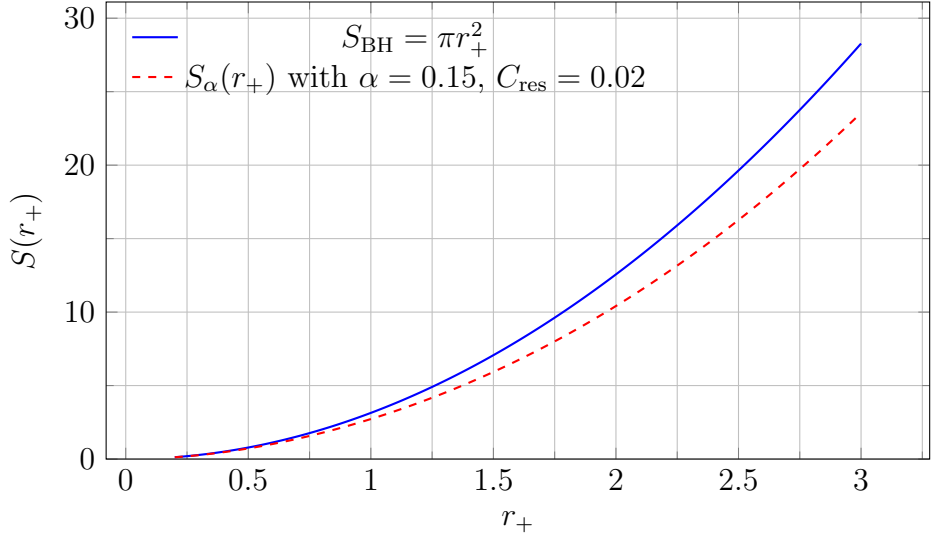

Figure~\ref{fig:entropyplot} provides a simple \texttt{pgfplots} implementation of the entropy profile. The plot compares the standard area law with the first-order residue-corrected entropy in the quadratic model.

For convenience, the corresponding standalone plotting core is
\begin{equation}
S_{\alpha}(x)=\pi x^{2}-\frac{8\pi^{2}C_{\rm res}}{3\alpha}x
+\frac{8\pi^{2}C_{\rm res}}{3\alpha\sqrt{3\alpha}}\arctan\!\left(\sqrt{3\alpha}\,x\right),
\end{equation}
with \(x\equiv r_{+}\).

\section{Conclusion}
\label{sec:conclusion}

Using the uploaded RVB-residue temperature prescription as input, we have constructed a black-hole entropy formula for \(f(Q)\) gravity from the first law of thermodynamics. The key result is the general integral expression
\begin{equation}
S(r_{+}) = \int^{r_{+}} \frac{2\pi u\,\Xi(u)}{\Xi(u)+4\pi C_{\rm res}u}\,\dd u + S_{0},
\end{equation}
where \(\Xi(u)=1+\psi(u)+u\psi'(u)\). For the quadratic model \(f(Q)=Q+\alpha Q^{2}\), this leads to the first-order closed entropy formula
\begin{equation}
S_{\alpha}(r_{+})
= \pi r_{+}^{2}
-\frac{8\pi^{2}C_{\rm res}}{3\alpha}r_{+}
+\frac{8\pi^{2}C_{\rm res}}{3\alpha\sqrt{3\alpha}}
\arctan\!\left(\sqrt{3\alpha}\,r_{+}\right)
+\cO(C_{\rm res}^{2}).
\end{equation}
The area law is recovered when the residue vanishes. Therefore the contour contribution enters entropy in a controlled and explicit way, exactly as one would expect from a residue-induced deformation of black hole thermodynamics.

A natural next step would be to compare this first-law entropy branch with a direct Noether-charge/Wald derivation in explicit \(f(Q)\) black hole solutions. That comparison would clarify whether the residue correction should be interpreted as a genuine modification of the horizon entropy, as an effective temperature renormalization, or as a combined thermodynamic reparameterization.

\bibliographystyle{unsrt}
\bibliography{rvb_fQ_entropy_refs}

@article{Bekenstein1973,
  author  = {Bekenstein, Jacob D.},
  title   = {Black Holes and Entropy},
  journal = {Physical Review D},
  volume  = {7},
  number  = {8},
  pages   = {2333--2346},
  year    = {1973},
  doi     = {10.1103/PhysRevD.7.2333}
}

@article{Hawking1975,
  author  = {Hawking, S. W.},
  title   = {Particle Creation by Black Holes},
  journal = {Communications in Mathematical Physics},
  volume  = {43},
  number  = {3},
  pages   = {199--220},
  year    = {1975},
  doi     = {10.1007/BF02345020}
}

@article{Wald1993,
  author  = {Wald, Robert M.},
  title   = {Black Hole Entropy is the Noether Charge},
  journal = {Physical Review D},
  volume  = {48},
  number  = {8},
  pages   = {R3427--R3431},
  year    = {1993},
  doi     = {10.1103/PhysRevD.48.R3427}
}

@article{RobsonVillariBiancalana2019,
  author  = {Robson, Charles W. and Villari, Leone Di Mauro and Biancalana, Fabio},
  title   = {Topological Nature of the Hawking Temperature of Black Holes},
  journal = {Physical Review D},
  volume  = {99},
  number  = {4},
  pages   = {044042},
  year    = {2019},
  doi     = {10.1103/PhysRevD.99.044042}
}

@article{Heisenberg2024,
  author  = {Heisenberg, Lavinia},
  title   = {Review on \(f(Q)\) Gravity},
  journal = {Physics Reports},
  volume  = {1066},
  pages   = {1--78},
  year    = {2024},
  doi     = {10.1016/j.physrep.2024.02.001}
}

@article{MandalWangSahoo2020,
  author  = {Mandal, Sanjay and Wang, Deng and Sahoo, P. K.},
  title   = {Cosmography in \(f(Q)\) Gravity},
  journal = {Physical Review D},
  volume  = {102},
  number  = {12},
  pages   = {124029},
  year    = {2020},
  doi     = {10.1103/PhysRevD.102.124029}
}

@article{HeisenbergKuhnWalleghem2022,
  author  = {Heisenberg, Lavinia and Kuhn, Simon and Walleghem, Laurens},
  title   = {Wald's Entropy in Coincident General Relativity},
  journal = {Classical and Quantum Gravity},
  volume  = {39},
  number  = {23},
  pages   = {235002},
  year    = {2022},
  doi     = {10.1088/1361-6382/ac987d}
}

@article{HammadEtAl2019,
  author  = {Hammad, F. and Dijamco, D. and Torres-Rivas, A. and B{\'e}rub{\'e}, D.},
  title   = {Noether Charge and Black Hole Entropy in Teleparallel Gravity},
  journal = {Physical Review D},
  volume  = {100},
  number  = {12},
  pages   = {124040},
  year    = {2019},
  doi     = {10.1103/PhysRevD.100.124040}
}

@article{GomesJimenezKoivisto2023,
  author  = {Gomes, D{\'e}bora Aguiar and Beltr{\'a}n Jim{\'e}nez, Jose and Koivisto, Tomi S.},
  title   = {Energy and Entropy in the Geometrical Trinity of Gravity},
  journal = {Physical Review D},
  volume  = {107},
  number  = {2},
  pages   = {024044},
  year    = {2023},
  doi     = {10.1103/PhysRevD.107.024044}
}

@article{Chen2025,
  author  = {Chen, Wen-Xiang},
  title   = {Calculating the Hawking Temperature of Black Holes in \(f(Q)\) Gravity Using the RVB Method: A Residue-Based Approach},
  journal = {Canadian Journal of Physics},
  volume  = {103},
  pages   = {531--542},
  year    = {2025},
  doi     = {10.1139/cjp-2024-0214}
}

\end{document}